%%%%%%%%%%%%%%%%%%%%%%%%%%%%%%%%%%%%%%
\input harvmac.tex

%\input  vshapka
%\def\subsubsec#1{$\underline{\rm #1}$}

%%%%%%%%%%%%%%%%%%%%%  Rublenye bukvy   %%%%%%%%%%%%%%%%%%%%%%%%
\def\IB{\relax\hbox{$\inbar\kern-.3em{\rm B}$}}
\def\IC{\relax\hbox{$\inbar\kern-.3em{\rm C}$}}
\def\ID{\relax\hbox{$\inbar\kern-.3em{\rm D}$}}
\def\IE{\relax\hbox{$\inbar\kern-.3em{\rm E}$}}
\def\IF{\relax\hbox{$\inbar\kern-.3em{\rm F}$}}
\def\IG{\relax\hbox{$\inbar\kern-.3em{\rm G}$}}
\def\IGa{\relax\hbox{${\rm I}\kern-.18em\Gamma$}}
\def\IH{\relax{\rm I\kern-.18em H}}
\def\IK{\relax{\rm I\kern-.18em K}}
\def\IL{\relax{\rm I\kern-.18em L}}
\def\IP{\relax{\rm I\kern-.18em P}}
\def\IR{\relax{\rm I\kern-.18em R}}
\def\IZ{\relax\ifmmode\mathchoice
{\hbox{\cmss Z\kern-.4em Z}}{\hbox{\cmss Z\kern-.4em Z}}
{\lower.9pt\hbox{\cmsss Z\kern-.4em Z}}
{\lower1.2pt\hbox{\cmsss Z\kern-.4em Z}}\else{\cmss Z\kern-.4em Z}\fi}
%%%%%%%%%%%%%%%%%%%% Calligraphic letters  %%%%%%%%%%%%%%%%%%%%%%%

%%%%%%%%%%%%%%%%%%%%%%%%%% Derivatives  %%%%%%%%%%%%%%%%%%%%%%%%

%%%%%%%%%%%%%%%%%%%% letters with bar %%%%%%%%%%%%%%%%%%%%%%%%%%

%%%%%%%%%%%%%%%%%%%%%%%%%%% Math symbols %%%%%%%%%%%%%%%%%%%%%%%

\def\Tr{\rm Tr}

%%%%%%%%%%%%%%%%%%%%%%%%%%%%%%%%%%%%%%%%%%%%%%%%%%%%%%%%%%%%%%%%%

% Something to deal with sub-sub-sections

\def\unlockat{\catcode`\@=11}
\def\lockat{\catcode`\@=12}

\unlockat
% Something to deal with sub-sub-sections

\def\newsec#1{\global\advance\secno by1\message{(\the\secno. #1)}
\global\subsecno=0\global\subsubsecno=0\eqnres@t\noindent
{\bf\the\secno. #1}
\writetoca{{\secsym} {#1}}\par\nobreak\medskip\nobreak}
\global\newcount\subsecno \global\subsecno=0
\def\subsec#1{\global\advance\subsecno
by1\message{(\secsym\the\subsecno. #1)}
\ifnum\lastpenalty>9000\else\bigbreak\fi\global\subsubsecno=0
\noindent{\it\secsym\the\subsecno. #1}
\writetoca{\string\quad {\secsym\the\subsecno.} {#1}}
\par\nobreak\medskip\nobreak}
\global\newcount\subsubsecno \global\subsubsecno=0
\def\subsubsec#1{\global\advance\subsubsecno by1
\message{(\secsym\the\subsecno.\the\subsubsecno. #1)}
\ifnum\lastpenalty>9000\else\bigbreak\fi
\noindent\quad{\secsym\the\subsecno.\the\subsubsecno.}{#1}
\writetoca{\string\qquad{\secsym\the\subsecno.\the\subsubsecno.}{#1}}
\par\nobreak\medskip\nobreak}

\def\subsubseclab#1{\DefWarn#1\xdef
#1{\noexpand\hyperref{}{subsubsection}%
{\secsym\the\subsecno.\the\subsubsecno}%
{\secsym\the\subsecno.\the\subsubsecno}}%
\writedef{#1\leftbracket#1}\wrlabeL{#1=#1}}% Macros for boxes
\lockat

\def\IL{\relax{\rm I\kern-.18em L}}
\def\IH{\relax{\rm I\kern-.18em H}}
\def\IR{\relax{\rm I\kern-.18em R}}
%why???\font\manual=manfnt
\def\dbend{\lower3.5pt\hbox{\manual\char127}}

\def\c{\cdot}
\def\IZ{\relax\ifmmode\mathchoice
{\hbox{\cmss Z\kern-.4em Z}}{\hbox{\cmss Z\kern-.4em Z}}
{\lower.9pt\hbox{\cmsss Z\kern-.4em Z}}
{\lower1.2pt\hbox{\cmsss Z\kern-.4em Z}}\else{\cmss Z\kern-.4em
Z}\fi}

% more macros, alphabetically

\def\IZ{\relax\ifmmode\mathchoice
{\hbox{\cmss Z\kern-.4em Z}}{\hbox{\cmss Z\kern-.4em Z}}
{\lower.9pt\hbox{\cmsss Z\kern-.4em Z}}
{\lower1.2pt\hbox{\cmsss Z\kern-.4em Z}}\else{\cmss Z\kern-.4em
Z}\fi}
\def\IB{\relax{\rm I\kern-.18em B}}
\def\IC{{\relax\hbox{$\inbar\kern-.3em{\rm C}$}}}
\def\ID{\relax{\rm I\kern-.18em D}}
\def\IE{\relax{\rm I\kern-.18em E}}
\def\IF{\relax{\rm I\kern-.18em F}}
\def\IG{\relax\hbox{$\inbar\kern-.3em{\rm G}$}}
\def\IGa{\relax\hbox{${\rm I}\kern-.18em\Gamma$}}
\def\IH{\relax{\rm I\kern-.18em H}}
\def\II{\relax{\rm I\kern-.18em I}}
\def\IK{\relax{\rm I\kern-.18em K}}
\def\IP{\relax{\rm I\kern-.18em P}}

\def\lies{{\underline{\bf s}}}
\def\inbar{\,\vrule height1.5ex width.4pt depth0pt}

\font\cmss=cmss10 \font\cmsss=cmss10 at 7pt
\def\IR{\relax{\rm I\kern-.18em R}}

\def\Tr{\rm Tr}

% Macros for boxes

\def\boxit#1{\vbox{\hrule\hbox{\vrule\kern8pt
\vbox{\hbox{\kern8pt}\hbox{\vbox{#1}}\hbox{\kern8pt}}
\kern8pt\vrule}\hrule}}
\def\mathboxit#1{\vbox{\hrule\hbox{\vrule\kern8pt\vbox{\kern8pt
\hbox{$\displaystyle #1$}\kern8pt}\kern8pt\vrule}\hrule}}

%% ANOTHER SET OF MACROS

\def\inbar{\,\vrule height1.5ex width.4pt depth0pt}

\font\cmss=cmss10 \font\cmsss=cmss10 at 7pt
\def\IR{\relax{\rm I\kern-.18em R}}

\def\Tr{\rm Tr}

%REFERENCES
%

\def\hh{hep-th/}

\lref\simons{  J. Cheeger and J. Simons, {\it Differential Characters and
Geometric Invariants},
 Stony Brook Preprint, (1973), unpublished.}

\lref\cargese{ L.~Baulieu,
{Algebraic Quantization of Gauge Theories}, Lectures given
at the Carg\`ese Summer School on   ``Perspectives in
fields and particles'', 198, p1, Proc. ed. by J.L.
 Basdevant and M. Levy  (Plenum Press, New-York, 1983).}

 \lref\antifields{   L. Baulieu, M. Bellon, S. Ouvry and  C.
Wallet,   Phys.Letters   B252  (1990) 387;  M.  Bocchichio, Phys. Lett.     
B187     (1987) 322 and    B 192  (1987) 31; R.  Thorn,
  Nucl. Phys.   B257
(1987) 61. }

 \lref\thompson{  G. Thompson,  Annals Phys. 205 (1991) 130;
  J.M.F. Labastida and  M. Pernici, Phys. Lett. 212B  (1988) 56;
  D. Birmingham, M.Blau,  M. Rakowski and G. Thompson,  Phys. Rep. 209 (1991) 129.}

  \lref\tonin{ Tonin}

 \lref\seibergsix{  O. Aharony, M. Berkooz, N. Seiberg,  {\it Light-Cone
Description of (2,0) Superconformal Theories in Six Dimensions},
  hep-th/9712117\semi  O.  J.  Ganor, David R.  Morrison, N.  Seiberg,
 {\it
Branes, Calabi-Yau Spaces, and Toroidal Compactification of the N=1
Six-Dimensional $E_8$ Theory}, hep-th/9610251, Nucl. Phys.  B487 (1997)
93-127\semi
N.  Seiberg,
{\it Non-trivial Fixed Points of The Renormalization Group in Six
Dimensions},  hep-th/9609161, Phys. Lett.  B390 (1997) 169-171.
}

 \lref\wittensix{E.  Witten, {\it New  Gauge  Theories In Six Dimensions},
  hep-th/9710065. }

\lref\orlando{
 O. Alvarez, L. A. Ferreira and J. Sanchez Guillen,
 {\it  A New Approach to Integrable Theories in any Dimension},
hep-th/9710147.
}
\lref\bks{  L.~Baulieu, H.~Kanno and I.~Singer,
{\it Special Quantum Field Theories in Eight and Other Dimensions},  
hep-th/9704167, Talk given at
APCTP Winter School on Dualities in String Theory  (Sokcho, Korea),
February 24-28, 1997\semi
  L.~Baulieu, H.~Kanno and I.~Singer, {\it Cohomological Yang--Mills
Theory
in Eight Dimensions}, hep-th/9705127, to appear in Commun. Math. Phys. }

\lref\wittentopo { E.  Witten,  {\it  Topological Quantum Field Theory}, \hh9403195,
Commun.  Math. Phys.  {117} (1988) 353.  }

\lref\wittentwist { E.  Witten, {\it Supersymmetric Yang--Mills theory on a
four-manifold}, J.  Math.  Phys.  {35} (1994) 5101.}

\lref\west{  L.~Baulieu and P.~West,
{  \it Six-Dimensional TQFTs and Twisted Supersymmetry},
 hep-th/9805200\semi
L.~Baulieu and E. Rabinovici, {\it Self-Duality and New TQFTs for Forms,
 }
hep-th/9805122.}

\lref\bv{ I.A. Batalin  and V.A. Vilkowisky,    Phys. Rev.
D28  (1983) 2567\semi  M. Henneaux,  Phys. Rep.  126   (1985) 1;
M. Henneaux and C. Teitelboim, {\it Quantization of Gauge Systems,}
Princeton University Press,  Princeton (1992).}

\lref\bs { L. Baulieu and I. M. Singer, {\it Topological Yang--Mills
Symmetry}, Nucl. Phys. Proc. Suppl.
15B (1988) 12\semi  L. Baulieu, {\it On the Symmetries of Topological Quantum
Field Theories},   hep-th/ 9504015, Int. J. Mod. Phys. A10 (1995) 4483\semi  
R. Dijkgraaf and G. Moore,
 {\it Balanced Topological Field Theories},
hep-th/9608169,   Commun. Math. Phys. 185 (1997) 411.}

\lref\kyoto {  L. Baulieu,   {\it Field Antifield Duality, p-Form Gauge
Fields and Topological Quantum
Field Theories},     hep-th/9512026,  Nucl. Phys.  B478 (1996) 431.  }

\lref\sezgin {
 L. Baulieu, E. Bergshoeff and E. Sezgin,
 {\it
Open BRST Algebra, Ghost Unification and String Field Theory, }
Nucl. Phys.   B307  (1988) 348.  }

\lref\strings {  L.  Baulieu, M. B. Green and E. Rabinovici {\it A Unifying
Topological Action for Heterotic and  Type II Superstring  Theories},
hep-th/9606080, Phys.Lett. B386 (1996) 91\semi
{\it   Superstrings from   Theories with $N>1$ World Sheet Supersymmetry},
 hep-th/9611136, Nucl. Phys. B498 (1997). }

\lref\sourlas{  G. Parisi and N. Sourlas,
{\it Random Magnetic Fields, Supersymmetry and Negative Dimensions},  Phys.
Rev. Lett.  43 (1979) 744; Nucl.  Phys.  B206 (1982) 321.  }

\lref\SalamSezgin{A.  Salam  and  E.  Sezgin,
{\it Supergravities in diverse dimensions}, vol.  1, p. 119\semi
P.  Howe, G.  Sierra and P.  Townsend, Nucl Phys B221 (1983) 331.}

\lref\nekrasov{ A. Losev, G. Moore, N. Nekrasov, S. Shatashvili,
{\it
Four-Dimensional Avatars of Two-Dimensional RCFT},  hep-th/9509151,
Nucl.  Phys.  Proc.  Suppl.   46 (1996) 130\semi L.  Baulieu, A.  Losev, N.
Nekrasov  {\it Chern-Simons and Twisted Supersymmetry in Higher
Dimensions},  hep-th/9707174, to appear in Nucl.  Phys.  B.  }

\Title{ \vbox{\baselineskip12pt\hbox{hep-th/9808055}
\hbox{CERN-TH-98-258}
%\hbox{KCL-MTH-98-12}
\hbox{LPTHE-98-25 }}}
{\vbox{\centerline{  On Forms with Non-Abelian   Charges and Their
Dualities
 }}}
\medskip
\centerline{Laurent Baulieu   }
\vskip 0.1cm
\centerline{CERN,
Geneva, Switzerland}
\vskip 0.1cm
\centerline{and}
\centerline{LPTHE\foot{UMR-CNRS associ\'ee aux  Universit\'es Pierre et
Marie Curie (Paris VI) et Denis Diderot (Paris~VII)}, Paris, France.}

 \medskip
\vskip  1cm
\noindent
We describe forms with  non-Abelian    charges.
We avoid the use of theories  with flat curvatures
by working in the context of topological field
theory. We obtain TQFTs for
a form and its dual. We leave open the question of getting gauges in
which the form, or its dual, can be gauged away, in such way that
the model has two dual formulations.
We give the example of  charged
 two-forms in six dimensions.
%\draft

\Date{May  1998}

\def\e{\epsilon}
\def\demi{{1\over 2}}

\def\a{\alpha}
\def\b{\beta}
\def\c{\gamma}
\def\m{\mu}
\def\n{\nu}
\def\r{\rho}

\def\L{L}
\def\X{X}
\def\V{V}

\def\P{\Psi}
\def\F{\Phi}
\def\Ic{{\it I}_{cl}}
\def\d{\delta}
\newsec{Introduction}
In this paper  we show that non-Abelian interactions can be
introduced   rather straightforwardly for topological quantum field
theories
 (TQFTs) of forms.

The natural generalization of the Abelian  gauge symmetries for  a
$p$-form $ B_p$ valued in some representation of a Lie algebra ${\it
{G}}$ implies the  introduction  a Yang--Mills field $A$ together with
$B_p$.
One   considers, as a first attempt, infinitesimal gauge transformations
  \eqn\gauge{\eqalign{
&\d A=  d\e +[A,\e]
\cr
&
\d B_p=  d\eta_{p-1} +[A,\eta_{p-1}]+[B_p,\e].
}  }
Here, $\e$ is the $0$-form parameter for the Yang--Mills
symmetry, and $\eta_{p-1}$ is an infinitesimal $(p-1)$-form, valued in
the same representation of ${\it {G}}$ as   $B_p$.

It is only when the Yang--Mills curvature $F_A=dA+\demi [A,A]$ vanishes
that the system of gauge transformations \gauge\ closes. Moreover, the
curvature $G_{p+1} =dB_p + [A, B_p ] $ does not transform covariantly,
except if $F=0$. This  makes  it  difficult to construct    an   action
invariant under the unrestricted set of transformations
 \gauge.

 Some of the  difficulties  triggered by the non-closure of gauge
transformations can be   overcome  by using the Batalin--Vilkoviski
formalism \bv. Indeed, the latter  is appropriate  for gauge symmetries
with infinitesimal gauge transformations, which only close modulo some
equations of motion. It was
  already noticed  in the past that, when one  quantizes
charged-forms, all relevant fields and antifields     fit in a
unifying  formalism, which generalizes   that of the  genuine Yang--Mills
case  (see e.g. \sezgin\ and
\kyoto).  However, the     invariant classical actions found in \kyoto\
were first order, and   describe    fields with vanishing  curvatures.
The situation was thus not  quite satisfying, since one  misses
Lagrangians     with squared curvature, that is, Lagrangians of the
Yang--Mills-type.
The fact that antifields and fields can be unified into  objects which
mix positive and negative ghost numbers, in a way that    closely
fits   the idea
  of duality, was however quite encouraging.

Another difficulty that one also  encounters with charged forms
 and   their gauge transformations
is that of their mathematical  definitions.
 There is, however, a
proposition that can be found in \orlando, which  could be a hint to define non
Abelian-forms and Yang--Mills fields in a unified
geometrical framework. Following eq.
(6.96) of \orlando, one can saturates the ``colour'' indices of the forms $B^i_p$
by matrices $S_i$ and    extend the    Lie algebra ${\it {G}}$ for the
Yang--Mills field $A=T^a A^a$ as follows:
\eqn\suggestion{\eqalign{
&  [T_a,T_b]=f_{ab}^c T_c, \cr
&[S_i, T_a]= r_{ia}^k S_k  ,\cr
&[S_i, S_j]= 0.
}  } Then $B_p=B^i_p S_i$ and one  defines the  curvature $ G_{p+1}
=dB_p+[A,P_b]=G_{p+1}^i S_i$.

The idea that we develop here is that, if one allows general transformations for
the $p$-form, while keeping track of the gauge symmetry \gauge\ by considering
the equivariant cohomology with respect to \gauge, we escapes almost by
definition all problems raised by the non-closure of the transformations
\gauge.  The natural framework for this is TQFT.
Thus, instead of \gauge, we  consider the following system of infinitesimal
transformations:
 \eqn\gaugetop{\eqalign{
&
\d B=  d\eta_{p-1} +[A,\eta_{p-1}]+[B_p,\e]+ \e_p,
}  }
where the $p$-form parameter $\e_p$ describes an infinitesimal
arbitrary transformation for $B_p$. Indeed,  arbitrary gauge
transformations build a closed algebra. Thus,  the presence of the
parameter
$\e_p$ in the gauge transformations allows for a  compensation for that
part of gauge transformations that  necessitates   constraints
to reach  a closed algebra. By introducing an equivariant cohomology in
the context of TQFT, it   becomes  possible  to consistently isolate
from the general transformations the  reduced set of gauge
transformations \gauge. One actually expects TQFTs whose actions are
supersymmetric  and contain  squared curvatures for the forms.

The   concept of   degenerate gauge symmetries, which characterize
TQFTs, must be   applied to these gauge transformations:  the number of
parameters
$\eta_{p-1}$ and $\e_p$ for the symmetry defined by    \gaugetop\
exceeds the number of degrees of freedom of   $B_p$. One can solve this
difficulty, and  distinguish between the  gauge transformations with
parameters $\eta_{p-1}$ or
$\e$ and the rest of the general transformations for  $B_p$.  This is an
almost obvious generalization of  the case of topological Yang--Mills
symmetry \bs.

It is actually simple to deduce from
\gaugetop\   a  BRST operator which is  nilpotent, independently  of any
constraint, and which consistently  separates between the general gauge
transformations and those tentatively  defined by \gauge. The existence
of  such an operator eventually leads  one  to the existence of an
invariant TQFT action, provided  topological gauge functions exist for
$B_p$.

 For the sake of notational simplicity, we first consider  the case of
two-form gauge fields, $p=2$.     Moreover, as for the   construction
of  a TQFT action, we   work  in  six dimensions  as in the Abelian case
analized in  \west. Afterwards, we   generalize our formulae.
For the general case, we introduce antifields,
in a Batalin--Vilkoviski approach, which provides an interesting
unification between all ingredients. We will consider
the definition of a TQFT for a $p$-form and a $D-p-2$-form
in $D$ dimensions. There is a natural  question about knowing wether
topological gauge functions exist such that the TQFT can be expressed
solely in terms of the $p$-form or  of the  ($D-p-2$)-form, with duality
transformations between the two formulations.  We leave it unexplored.

\newsec{The example of charged 2-forms in six dimensions}
 We  call
$B^1_1$ and
$B^2_0$ the one-form ghost and zero-form ghost  of ghost for  the gauge
symmetries of a charged two-form
$B_2$. Then,  $\Psi^1_2$,  $\Phi^2_1$ and $\Phi^3_0$ are the topological ghost and
ghosts of ghosts  of
$B_2$. All these fields are in the same representation of ${\it G}$ as
$B_2$.
 (Following the   conventional notation the upper index of a form will be its 
  ghost-number and the lower one its  ordinary form-degree.)

Let $D_A\cdot=d+[A,\cdot ]$ be the covariant
derivative with respect to $A$
and $F_A=D_A\wedge D_A=dA+A\wedge A$ be the curvature of $A$. We define the  
curvature of $B_2$ as $G_3=D_A B_2$.

The topological BRST symmetry associated to the gauge symmetry
\gaugetop\  is defined from the expansion in ghost number of the
following equations:
\eqn\brsa{\eqalign{
(s+d)(A+c)+\demi[A+c,A+c] =F_A,
}}
 \eqn\brsb{\eqalign{
(s+d)(B_2+B_1^1+B_0^2)+ [A+c,B_2+B_1^1+B_0^2] =G_3+\P_2^1+\F_1^2+\F_0^3.
}}
($c$ is the Faddeev--Popov ghost  of $A$.) These equations   define  the  
action of $s$, with  the desired property that
$s^2=0$, if they are comparable with   the Bianchi identity $D_AD_A=F_A$ and  
$D_A F_A=0$. Thus,  we must
define the action of $s+d$ on the fields of the right-hand side of \brsb\
as:
 \eqn\brsbt{\eqalign{&
(s+d)(G_3+\P_2^1+\F_1^2+\F_0^3)+ [A+c,G_3+\P_2^1+\F_1^2+\F_0^3]
=[F_A,B_2+B_1^1+B_0^2]. \cr  }}
This warranties that $(s+d)^2=0$, and thus $s^2=0$ on all fields.

One might be interested in a  detailed expression of $s$ on the fields.
It  is obtained by expanding the BRST equations \brsb\ and \brsbt. One finds  
the following
expression for the action of $s$:  \def\lies{s}
\eqn\syms{\eqalign{& \lies B_2 = \P^1_2
 -
D_A B_1^1-[c,B_2]
\cr
& \lies B_1^1 = \F^2_1
 -
D_A B_0^2-[c,B_1^1]
\cr
&   \lies B_0^2 = \F^3_0
 -[c, B_0^2]
}  }
\eqn\symst{\eqalign{  &
\lies \P^1_2=-D_A \F^2_1
  +[F_A, B_0^2]-[c,\P^1_2]
\cr
&
\lies \F^2_1=-D_A \F^3_0
 -[c,\F^2_1]
\cr &
\lies \F^3_0=
 -[c, \F^3_0].
\cr
}  }
Of course, we have $sA=-D_Ac$ and $sc=-\demi[c,c]$.  No antifield
dependence is necessary, since the general transformations described by $s$  
determine a
closed algebra.  However,
  we will see in  the next section how antifields can be
consistently introduced in the BRST equations. We can already guess
  in a very elementary  (although unnatural) way
that antifields can be introduced by brute force in \syms\ and
\symst,  by redefinitions of topological ghosts.

As for writing an action that  is  a classical topological invariant, and thus a
starting point for a TQFT action ready for path integration,  we need to specify
the dimension of space. In six  dimensions, we need a     pair of two-forms,
$B_2$ and
$^cB_2$, because
$G_3$ has an odd degree (see
\west\ for the Abelian situation). The following action is a possibility,  which
generalizes  the Abelian topological term
$\Ic=\int_6 dB_2\wedge d^cB_2$ in \west:
 \eqn\Iclnab{\eqalign{\Ic = \int _6 {\Tr} \ (\ DB_2\wedge D{^cB}_2 +F\wedge
[{^cB}_2,{ B}_2]\ ). }}
The doubling of the number of forms   amounts to have  mirror  equations as in 
\brsb\ and \brsbt, with a duplication of all ghosts. (An index
$^c$ is   introduced  for all mirror fields.)

Using the techniques detailed in \west\ (the only difference is that we must
define the
antighosts and Lagrange multipliers as elements of the same Lie algebra
representation as $B_2$), one can impose in a    BRST way  the following
  choices of gauge functions for $B_2$ and $^cB_2$:
\eqn\motion{ D_{A [\m}
B_{\nu\rho]}+\e_{\mu\nu\rho\a\b\c}
D_{A}^{[\a}{^cB}^{\b\c]}
}
and for the   ghosts and antighosts:
\eqn\clever{\eqalign{ &
D_{A}^{\n}\Psi^1_{[\m\n]}; \quad   D_{A}^{\m}\Phi^2_\m; \quad D_{A}^{\m}
\Phi_\m^{-2}
\cr &
D_{A}^{\n}{^c\Psi}^1_{[\m\n]}; \quad   D_{A}^{\m}{^c\Phi}^2_\m; \quad D_{A}^{\m}
{^c\Phi}_\m^{-2}. } }

 The   six-dimensional BRST invariant   action which uses  the gauge
functions \motion\ and
\clever\  is:
 \eqn\symQfc{\eqalign{ \int _6 &
 |D_{A [\m} B_{\nu\rho]}|^2
+ \chi^{\m\n\r}
D_{A [\m} \Psi^1_{\nu\rho]}
+\eta_\m ^{-1}D_{A  }^\n\Psi_{[\m\n]}^{1}
\cr &
+\X^1D_{A }^{ \m}\eta_\m^{-1}
+  \Phi^{-3}D_{A  \m}D_{A  }^\m \Phi^{ 3}
\cr &
+  \Phi^{-2}_\m D_{A  \m}D_{A}^{\m}\Phi^{ 2}_{\n ]}
+D_{A  }^{\n}   \Phi_\n^{-2}D_A^\m \Phi^{ 2}_\m
\cr &
-{\rm{the \ same \ expression \  with \ all \ fields \ \phi\  replaced\  by\   
^c\phi}}.
\cr
}  }
 This action, which is an $s$-exact term, is  the non-Abelian extension of  
that constructed
in \west.    It possesses a   genuine Yang--Mills invariance
since all gauge functions in
\motion\ and
\clever\ are gauge-covariant under   Yang--Mills transformations. The  
gauge-fixing of this symmetry is quite an obvious
task. We   can for instance   go in a BRST invariant way to a
Feynman--Landau-type gauge for
$A$.   We have  also to gauge-fix the ordinary
gauge invariance of the two-forms $B_2$ and ${^c B}_2$.  We must introduce  
the ordinary ghosts and ghosts of ghosts of $B_2$ and ${^c B}_2$,
and proceed  by replacing all derivatives by covariant ones in the gauge  
functions of   the Abelian case given in \west. This  can be done  
straightforwardly, and it gives additional ghost contributions to \symQfc.
In the next section,  the ordinary ghosts of the two-forms will be introduced  
with a wider perspective.

Let us now  stress that  it is possible to  couple  the theory to a
  Yang--Mills TQFT instead to the genuine Yang--Mills theory.
To do so, $F_A$     has to be
replaced  whereas it appears in
\brsa,
\brsb\ and \brsbt\  by
$F_A+\P^1_1+\F^2_0$. Here, $ \P^1_1 $ and $ \F^2_0$ are the
 topological ghost and ghost of ghost of the Yang--Mills theory  
\wittentopo\bs, and thus \brsa\ becomes:
\eqn\brsat{\eqalign{
(s+d)(A+c)+\demi[A+c,A+c] &=F_A+\P^1_1+\F^2_0 ,\cr
(s+d)(F_A+\P^1_1+\F^2_0)+ [A+c,F_A+\P^1_1+\F^2_0] &= 0.
}}
Then,  $sA=\P^1_1-D_Ac$ and
$sc=\F^2_0-\demi[c,c]$, and eqs.
\syms\ and
\symst\ are  modified accordingly.

Since  the action
\Iclnab\ is a  topological term, it is still invariant  when  the
Yang--Mills field transforms as   in \brsat.  The   gauge fixing for the
two-form gauge field   goes the same  way  as above, with the same gauge
functions  as in  \motion\ and
\clever. However, new   contributions come  by varying   the gauge field
$A$, which is present in these gauge functions. This   gives   couplings
between the topological ghost $\P^1_1$ of $A$ and the rest of the fields
in  \symQfc.

As for the dynamics of the Yang--Mills field,  one can add to \Iclnab\
an action $\int {\Tr} F_{\m\n}F^{\m\n}$ when  the gauge symmetry of $A$
is defined by \brsa. On the other hand,  when it is defined by \brsat,
one can only add to \Iclnab\ a topological  Yang--Mills gauge action,
as, for instance,  the one defined in \bks\ for the six-dimensional
case.  There is another option (see next section),  which consists in
introducing a TQFT with an additional three-form, starting from    the
topological term
$\int_6 {\Tr} F_A\wedge D_AZ_3$. This   provides  an additional  TQFT
action  of the Bogomolny type.

The important result is the existence of the action  \symQfc. It shows
that we can define an invariant action depending on   charged two-forms,
with squared curvatures. Because it possesses a BRST invariance of the
same type as the one of the Abelian case, its tree approximation is also
related to Poincar\'e supersymmetry as in \west.

In order to look for  observables of the TQFT,   one remarks that the
simplest possibility come from the  cocycles that are defined by
the ghost expansion of:
 \eqn\cocycle{\eqalign{ \Delta_6=
\Tr \ \big (   (D_AB_2+ \P_2^1+\F_1^2+\F_0^3)\wedge(
D_A{^cB}_2+{^c\P}_2^1+{^c\F}_1^2+{^c\F}_0^3) \cr
+F_A\wedge [{^cB}_2+{^cB}^1_1+{^cB}^2_0
,B_2+B^1_1+B^2_0]\ \big ). }}

 One has by   construction that
$(s+d)\Delta_6=0$, which implies   $ s\int_{\Gamma_d} \Delta
_d^{6-d}=0$, where
$0\leq d\leq 6$, and the integration is done over a $d$-cycle
${\Gamma_d}$. The $\int_{\Gamma_d} \Delta _d^{6-d}$ are thus the
candidates as   observables of the TQFT. This  generalizes the
Yang--Mills situation \wittentopo\bs. However, we are not yet in
a position
to understand the meaning of these observables. It could be that they
should only be considered for  a gauge field $A$  with a vanishing
curvature.

\newsec{Generalization: TQFTs for dual pairs}

 We now wish  to reach a more detailed  understanding of the BRST
structure introduced in
\brsb\ and  \brsbt, and \brsa\ or \brsat. For this,   the
antifield-formalism of Batalin--Vilkoviski \bv\    turns out to be
useful, not only for solving the questions related to the non-closure,
but also for enlightening the idea of duality \kyoto. Introducing
antifields is natural  in various   situations where the closure of gauge
transformations holds only up to equations of motion (for example, see
refs. \antifields).  It is also known that  the  four-dimensional
Yang--Mills topological theory can  be   related    to  a  non-Abelian
$B-F$ system   (see e.g.
\thompson). So, even in this   simpler   case,    the use  of antifields
sheds a particular  light on the theory.

In what follows,  we  introduce  the relevant fields and antifields  for
quantizing \Iclnab\ and construct a  Batalin--Vilkoviski action. We will
obtain  at once    the topological term  \Iclnab\ and   the
complete TQFT action  with its antifield dependence. The    latter
defines the    BRST symmetry equations by a  master equation.

We can   consider a more  general case than  that   we have  described
above. Instead of $B_2$ and
${^c B}_2$ in six dimensions, we   introduce a
   $p$-form $U_p$ and   a $(D-p)$-form $V _{D-p}$ in $D$
dimensions with the same charges under the Yang--Mills symmetry. We
associate  new   fields  to
 $U_p$,
 $V_{D-p-2}$ and     the Yang--Mills field  $A$: they  are   respectively
 a      $(D-p-1)$-form $X_{D-p-1}$,  a $(p+1)$-form $Y _{p+1}$,   and a
 $(D-2)$-form $B _{D-2}$.   We also introduce a
$2$-form $W _{ 2}$ with its companion, a      $(D-3)$-form $Z_{D-3}$.
The forms  $A$, $Z_{D-3}$, $B _{D-2}$, and $W$   are valued  in  ${\it
G}$. The other  fields are valued in a  given representation of ${\it
G}$ that  can be chosen at will.

The TQFT that will be constructed will depend  dynamically  on the pair
of fields
 $U_p$ and
 $V_{D-p-2}$. It is a very intriguing question to ask whether
there exists
different topological gauge functions   that  define
TQFTs,   which can be separately
expressed solely in terms of  $U_p$  or of
 $V_{D-p-2}$  (and supersymmetric partners), in such a way that
the two formulations  can be transformed into each other by
duality transformations.

The justification for introducing all the other fields that accompany
$A$, $U_p$ and
 $V_{D-p-2}$  can be found in
\kyoto. Let us summarize the argument. One introduces
``generalized forms''  for which  the  degree   is defined as    the sum
of the   ghost number and ordinary form degree. Thus, a
``generalized $p$-form'' is made of a sum of forms, each of them having
an ordinary form degree and an    integer  value  for the ghost
number, positive or negative, such that  the  sum  of its   ordinary
degree and ghost number   is equal to
$p$. The ordinary form degree can  only run from $0$ to $D$. It follows
that  the expansion in ghost number of a ``generalized $p$-form'' always
admits a finite number   of terms, equal to $D+1$, independently of the
value of $p$. Among these $D+1$ independent  terms, there are  forms with a
positive ghost number and forms with a negative one. The former
are ordinary ghosts, which can be interpreted as ghosts or ghosts of
ghosts associated with
 the parameters of the (degenerate) gauge symmetry.
By definition, the components  with negative ghost numbers  are
antifields.  All these fields are supposed to play a role in the TQFT.
  Since the    ghost number of the antifield of a field with ghost
number $g$ is equal to $-g-1$ \bv, the   antifield  of a form $\phi^{g
}_{q }$ in $D$ dimensions is a  form $\psi^{-g-1}_{D-q }$. Thus, one
  can     unify all the  ghosts fields and the   antifields
 for the  quantization of  a pair of classical forms $ \phi_q $ and   $
\psi _{D-q-1}$  as the components of    ``generalized forms''
$ \tilde \phi_q $  and $ \tilde \psi _{D-q-1}$, with degrees $q$ and
${D-q-1}$  respectively.
  The component   $\phi^{-g-1}_{q+g+1}$ in the expansion of
$\tilde \phi_q $ can be identified, for $g\geq 0$, as  the antifield  of
the  ghost
$\psi^{g}_{D-q-g-1}$   in the expansion  of  $\tilde \psi _{D-q-1}$.

These remarks justify
 the introduction of all   ``companion''  fields
mentioned above for writing an action for the forms $A$, $U_p$
and
$V_{D-p-2}$. Let us now write in more detail  the    ghost expansions
of all these fields.  For a better understanding, we write,   in each
pair of  the  following equations, fields and antifields  on top
of each other:
\def\A{{\tilde A}}
\def\B{{\tilde B}}
\def\U{{\tilde U}}
\def\V{{\tilde V}}
\def\X{{\tilde X}}
\def\Y{{\tilde Y}}
\def\W{{\tilde W}}
\def\Z{{\tilde Z}}

 \eqn\expA{\eqalign{\matrix{     &
{  \A } &=&
{    c }  &+&
{    A}  &+&
{    B_{  2 }^{  -1 } }  &+&
{  B_{  3 }^{  -2 }  }  &+&
{  \dots }  &+&
{   B_{  3 }^{  -2 }}
\cr
         &
{  \B_{D-2} } &=&
{    c_D^{-2} }  &+&
{    A_{D-1}^{-1}}  &+&
{   B_{  D-2 }^{    } }  &+&
{   B_{  D-3 }^{ 1 }  }  &+&
{ \dots  } &+&
{   B_{  0}^{  D-2 }}  }}}

\eqn\expW{\eqalign{\matrix{&
{  \W_2 }  &=
{   \F_0^2} &+&
{     \P_{  1 }^{  1 }}  &+&
{    W_{  2 }^{  } }  &+&
{  Z_{  3 }^{  -1}  }  &+&
{  Z_{  4 }^{  -2}  }  &+&
{  \ldots}  &+&
{   Z_{ D}^{  -D+2 }}
\cr
 &
{  \Z_{D-3} }  &=
{ \F_{  D }^{ -3 } } &+&
{     \P_{  D-1 }^{ -2 }} &+&
{   W_{  D-2 }^{-1    } } &+&
{   Z_{  D-3 }^{   }  } &+&
{  Z_{  D-4 }^{  1 } }  &+&
{   \ldots}  &+&
{    Z_{ 0 }^{  D-3 }}
\cr
&
  }}}

\eqn\expU{\eqalign{\matrix{&
{  \U_p }  &=&
{   U_0^p}+
{    \ldots} +{    U_{  p-1 }^{  1} }  +&
{  U_{  p  }^{  }  }  &+&
{ X_{  p-1}^{ -1}  }  &+&
{  \ldots}&+
{   X_{ D}^{  -D+p }}
\cr
 &
{  \X_{D-p-1} }   &=&
{U_{  D }^{ -p-1 } }+
{    \ldots} +
{  U_{    D-p+1 }^{-2    } } +&
{   U_{  D-p }^{ -1  }  } &+&
{  X_{  D-p-1}^{   } }  &+&
{   \ldots} &+
{   X_{ 0 }^{  D-p-1 }}
\cr
&
  }}}

\eqn\expV{\eqalign{\matrix{&
{  \V_{D-p-2} }  &=&
{   V_0^{D-p-2}}+
{    \ldots}
 +{    V_{ D-p-3} ^{  1}}   +&
{  V_{  D-p-2} ^{  } }   +&
{ Y_{ D- p-1}^{ -1}  }   +&
{ {    Y_{  D-p }^{  -2 }+...}} +
{   Y_{ D}^{  -p-2 }}
\cr
 &
{  \Y_{p+1} }   &=&
{V_{  D }^{ -D-p-1 } }+
{    \ldots} +
{  V_{   p+3 }^{-2    } }\ \ \  +&
{   V_{  p+2 }^{ -1  }  }\ \  \ \ +&
{  Y_{  p+1}^{   } } \ \ \   +&
{    Y_{  p }^{  1 }+... } +
{   Y_{ 0 }^{  p+1 }}
\cr
&
  }}}
We recognize some the topological ghosts introduced  in the particular
case of the six-forms. For instance, the expansion of $\W_2$ contains
the topological ghosts $\Psi^1_1$ and $\Phi ^2_0$ of  $ A$ and that of
$\Y_{p+1}$  contains the topological ghosts of  $ U_{p }$, which are
identified as $ Y_{  p }^{  1 }$,
$ Y_{  p -1}^{  2 }$, ..., $ Y_{  0 }^{  p+1 }$. The antifield
identification
  of the fields with negative ghost number   will become quite obvious
by looking at  the  BRST equations that we will  shortly obtain, by
imposing a
 Batalin--Vilkoviski type master equation.

We now introduce the TQFT Lagrangian density as the following $D$-form
with ghost number zero:
\def\w{\wedge}
\def\DA{{D_A}}
\def\DAT{{D_\A}}
\def\FA{{F_A}}
\def\FAT{{F_\A}}
 \def\L{{\cal L}}
 \def\I{{\cal I}}
\eqn\IUV{\eqalign{\L_D =   {\Tr} \ \Big(\ &
\X_{D-p-1} \w \Y_{p+1}\ +\
\X_{D-p-1} \w  \DAT \U_p \ +\
\Y_{p+1}  \w \DAT \V_{D-p-2}
 \cr &
+\B_{D-2} \w \W_2
+  \B_{D-2} \w \FAT
+ \W_2 \w
(\DAT \Z_{D-3} +[\U_p,\V_{D-p-2}]
\ )\ \Big)\
\Big| ^0_D\ . }}
This Lagrangian    is of  first order
and metric-independent. The intuition for writing \IUV\ is that, if one
considers its purely classical  part  $\L_{cl,D}$, by  setting all ghosts and  
antifields
equal to
zero,  (which means getting rid  of all
tildes in  \IUV), and if one  eliminates
$ X_{D-p-1}$,
$ Y_{p+1}$,
$B_{D-2}$ and
$Z_{D-3}$
by their algebraic equations of motion $Y=\DA U$, $X=\DA V$ and $W=\FA $, we
have:
\eqn\IUVcl{\eqalign{\L_{cl,D} \sim    {\Tr} \ \Big(\ &
  \DA   U_p   \w \DA   V_{D-p-2}
 +\FA \w [ U_p, V_{D-p-2}]\cr &+
\FA \w \DA  Z_{D-3}
\Big) . }}
As a result of the elimination of the fields
$\tilde X$ and  $\tilde Y$,
the Lagrangian density
\IUVcl\ is locally $d$-exact, contrarily to
\IUV.  The integral over $D$-dimensional space
of the density
\IUVcl\ can be considered as a topological term.  Notice that the  last
term  of
\IUVcl\  is    $d$-exact by itself.    In the
particular case studied in the  previous section, we  had
not   introduced such  a term   in
\Iclnab.  However, we   already observed
that we had the freedom to consider arbitrary redefinitions of $A$ in
\Iclnab, which is precisely the symmetry of $\int _D \FA \w \DA
Z_{D-3}$.

We define the   invariant Batalin--Vilkoviski-type action:
\eqn\IBV{\eqalign{\I_D=\int _D \L_D . }}
$\I_D $ satisfies a master equation, which is equivalent to its BRST symmetry:
\eqn\IBV{\eqalign{s \I_D=0,  \quad
s\phi = {{\delta \I_D} \over {\delta\psi}} ,  \quad
s\psi = {{\delta \I_D} \over {\delta\phi}}, }}
where, generically, $\psi$ is the antifield of $\phi$.

Thus, the BRST equations for all   fields and antifields defined in  eqs.  
\expA--\expV,  can be   obtained  by varying   \IUV. They are, in a compact  
form:
\eqn\brtsA{\eqalign{s\A&=-\FAT+\W_2\cr
s\W_2&=-\DAT\W_2 . }}
\eqn\brtsU{\eqalign{s\U{_p}&=-\DAT \U_{p}+\Y_{p+1}\cr
s\Y_{p+1}&=-\DAT\Y_{p+1}. }}
\eqn\brtsV{\eqalign{
s\V_{D-p-2}&=-\DAT \V_{D-p-2}+\X_{D-p-1}\cr
s\X_{D-p-1}&=-\DAT\X_{D-p-1}. }}
\eqn\brtsB{\eqalign{
s\Z_{D-3}&=-\DAT\Z_{D-3}+[\U_{p }, \V_{D-p-2}] +\B_{D-2}
\cr
s\B_{D-2}&=-\DAT \B_{D-2}
+[\X_{D-p-1},\U_p]
+[\Y_{p+1}, \V_{D-p-2}]
+[\W_{2}, \Z_{D-3}]
. }}
(In order to reach a more  detailed  expression of $s$ on all fields and
antifields,  one has to do a further  expansion in  ghost number,  as
one does   to obtain  \syms\ and \symst\ from \brsb\ and \brsbt.
For instance,
\brtsA\ gives $sA=\Psi^1_1 -Dc$, $sc=\Phi^2_0-\demi[c,c]$, etc...)

Equations    \brtsA--\brtsB\ are quite illuminating: if one sets all antifields
equal to zero, one recovers    field transformations similar to those
in \brsb,
\brsbt\ and \brsat;   eqs.  \brtsA--\brtsB\ contain more information than
eqs.   \brsb\ and \brsbt\ since they   encode the way the antifields
transform.  This gives at once the BRST transformation, and a field and  
antifield dependent action,
which we can  gauge-fix  by the Batalin--Vilkoviski method.

The transformations of
$\A$, $\U$, $\V$ and $\Z_{D-3}$ show that these fields have a symmetry
of the     topological type. Indeed, the occurrence of  $\W$, $\Y$,
$\X$ and $\B_{D-2}$ in  the right-hand-side of    \brtsA-\brtsB\
implies   a symmetry made of arbitrary shifts  defined modulo gauge
transformations. Moreover, the topological ghosts of $\A$, $\U$, $\V$
and $\Z_{D-3}$ are now identified as the ordinary ghosts and ghosts of
ghosts of   $W$, $Y$, $X$ and $B_{D-2}$. One can   see the way the gauge
symmetry is contained in the topological symmetry, by setting some of the  
topological ghosts  equal to
zero.

As already noted, the     classical fields
$W$, $Y$, $X$ and $B_{D-2}$ can be eliminated   from the action because of their
algebraic equations of motion. This operation  gives  the  topological classical
Lagrangian
\IUVcl, plus   antifield-dependent terms. By
applying the Batalin--Vilkoviski procedure, these antifield-dependent terms
eventually  determine   a fully gauge-fixed action, which is analogous to
\symQfc. However, one needs to guess
     relevant  gauge functions.
 The latter are  expected to be of the self-dual type,
 as, for instance,   those defined by \motion\ and
\clever, which are generically of the type
$D_A U =*(D_A V)+\ldots$  Moreover, one can investigate
the question of having more refined gauge functions, which would give a
theory depending only on
$U$ or $V$, with duality transformations between both formulations.

Let us conclude by the following remark. We can construct
   topological terms like \IUV.  They  are $D$-forms that can be written
locally as $d$-exact terms. In the $D$-dimensional theory, and after
the introduction of topological ghosts, they determines
cocycles, analogous to those defined, in a particular case,  in
\cocycle. These  cocycles are candidates for observables in potentially
interesting TQFTs.  They also generate, by the   descent equations for
forms \cargese, consistent anomalies for the ordinary gauge symmetries of
forms   in $(D-2)$-dimensional field theories.

\vskip 0.5cm

 \listrefs
\bye